\documentclass[11pt]{article}
\pdfoutput=1

\usepackage{aas_macros,amsmath,amssymb,comment,esint,graphicx,mathtools}
\usepackage[margin=.8in,letterpaper]{geometry}
\usepackage[colorlinks=true]{hyperref}
\usepackage[affil-it]{authblk}
\usepackage{url}
\usepackage{subcaption}
\usepackage{graphicx}
\usepackage[mathscr]{eucal}
\usepackage[normalem]{ulem}

\usepackage{epsf,amsmath,bbold,amsfonts,stmaryrd}

\usepackage[utf8]{inputenc}
\usepackage{hyperref}
\usepackage{cite}
\usepackage{mathrsfs}
\usepackage{appendix}
\usepackage{amssymb}
\usepackage{float}
\usepackage{color} 
\hypersetup{pageanchor=false}
\usepackage{indentfirst}
\usepackage{url}
\usepackage{float}
\usepackage{caption}
\usepackage[numbers,square,comma,sort&compress,merge]{natbib}
\usepackage{esint}
\usepackage{overpic}

\numberwithin{equation}{section}
\let\originalleft\left
\let\originalright\right
\renewcommand{\left}{\mathopen{}\mathclose\bgroup\originalleft}
\renewcommand{\right}{\aftergroup\egroup\originalright}
\mathcode`\*="8000
{\catcode`\*=\active\gdef*{\mathclose{}\,\mathopen{}}}

\newcommand{\be}{\begin{equation}}
\newcommand{\ee}{\end{equation}}
\newcommand{\bea}{\setlength\arraycolsep{2pt} \begin{eqnarray}}
\newcommand{\eea}{\end{eqnarray}}
\newcommand{\nn}{\nonumber}

\def \nn {\nonumber}

\begin{document}
\title{Circular orbits and polarized images of charged particles orbiting Kerr black hole with a weak magnetic field}

\author{Tsai-Chen Lee$^{1}$, Zezhou Hu$^{1}$, Minyong Guo$^{2\ast}$, Bin Chen$^{1,3,4}$}
\date{}

\maketitle

\vspace{-10mm}

\begin{center}
{\it
$^1$Department of Physics, Peking University, No.5 Yiheyuan Rd, Beijing
100871, P.R. China\\\vspace{4mm}

$^2$ Department of Physics, Beijing Normal University,
Beijing 100875, P. R. China\\\vspace{4mm}

$^3$Center for High Energy Physics, Peking University,
No.5 Yiheyuan Rd, Beijing 100871, P. R. China\\\vspace{4mm}

$^4$ Collaborative Innovation Center of Quantum Matter,
No.5 Yiheyuan Rd, Beijing 100871, P. R. China\\\vspace{2mm}
}
\end{center}

\vspace{8mm}

\begin{abstract}
In this work, we study the circular motions of charged particles and their polarized images around the Kerr black hole immersed in a weak magnetic field. We pay special attention to the case where the magnetic field and the charge-to-mass ratio are insignificant. Thus the effective potential along the radial motion reduces to a cubic form approximately so that we can express the radius of the innermost stable circular orbit analytically in terms of the energy and angular momentum of charged particles. Moreover, we computed the polarized radiations of these particles and obtained the polarized images semi-analytically for different spins, observational angles, and prograde and retrograde orbits. In particular, We find that these parameters significantly impact the polarization rotation and the magnitude of the polarization flux.
\end{abstract}

\vfill{\footnotesize $\ast$ Corresponding author: minyongguo@bnu.edu.cn}
\maketitle
\newpage
\section{Introduction}

Black holes are among the most fascinating predictions of general relativity, and accumulating evidence supports their existence in the universe. Recent detections of gravitational waves by LIGO and Virgo \cite{LIGOScientific:2016aoc, LIGOScientific:2016sjg, LIGOScientific:2017vwq, lipunov2017master} and imaging of black holes by the Event Horizon Telescope (EHT) collaboration \cite{EventHorizonTelescope:2019dse, EventHorizonTelescope:2019jan, EventHorizonTelescope:2019ths, EventHorizonTelescope:2019pgp, EventHorizonTelescope:2019ggy, EventHorizonTelescope:2022xnr} have increased our confidence in the accuracy of general relativity and the existence of black holes. General relativity and other theories of gravity predict various types of black holes, including neutral, charged, magnetic, or both charged and magnetic. Astrophysical black holes are typically expected to be neutral because they selectively accrete ambient matter. Despite this, these black holes can still generate a strong electromagnetic field that creates charged particles through pair production, as explained in the Blandford and Znajek \cite{Blandford:1977ds}. Furthermore, weakly charged black holes may exist in the universe, and one possible mechanism for charging them is the accretion of free charged particles to neutralize the electric field generated by a rotating black hole immersed in a magnetic field \cite{wald}.

The environment surrounding an astrophysical black hole is highly complex, containing plasma that generates magnetic fields and forms an accretion disk. Numerous observations have suggested the existence of magnetic fields outside black holes, including bright "flares," polarized broad H$\alpha$ lines, and polarization of accretion disk radiation near active galactic nuclei \cite{dovvciak2004polarization, Zajacek:2018vsj, Daly_2019,2010arXiv1002.4948P}. Recently, the EHT collaboration published a polarized image of the M87 black hole which provides further evidence of a magnetic field around the black hole \cite{EventHorizonTelescope:2021bee, EventHorizonTelescope:2021srq, narayan2021polarized, Himwich:2020msm}, as it is consistent with specific models of the magnetic field.

Studying the effects of the magnetic field around a black hole is crucial for interpreting information obtained from polarized images of the black hole. In this study, we are using a simplified model that disregards the interaction between charged particles and approximates their motions as point-like. We then examine the polarized radiations resulting from the accelerated motion of the charged particle in a magnetic field, approximating it as electromagnetic radiation from point particles. Especially when the charged particles are in relativistic motion, the resulting radiation is commonly referred to as synchrotron radiation. Moreover, we assume that charged particles are confined to move in circular paths on the equatorial plane beyond the innermost stable circular orbit (ISCO). Factors such as the influence of the Lorentz force on charged particles' ISCOs and the polarized image of radiations in the presence of magnetic fields are essential. The ISCOs of charged particles around a Schwarzschild and Kerr black hole immersed in a weak, axially symmetric magnetic field have been widely studied, with many exciting results found \cite{frolov, Zahrani:2022fdd, Shiose:2014bqa, AlZahrani:2014dfi}. For strong magnetic fields, spacetimes can be modeled by Schwarzschild and Kerr black holes that are embedded in the Melvin universe \cite{ernst, kerrmel}, and the corresponding ISCOs for massive particles have also been studied in the literature \cite{Lim:2015oha, Zhu:2022amy, Li:2018wtz}. In addition, the polarized image of radiations originating from circularly orbiting charged particles acted upon by Lorentz force has been investigated in Schwarzschild and Kerr black holes with weak or strong magnetic fields \cite{stab1289, Lupsasca:2018tpp, EventHorizonTelescope:2021btj, Gelles:2021kti, Qin:2021xvx, Zhang:2021hit, Hu:2022sej, Zhu:2022amy, Qin:2022kaf}.

In this work we revisit the ISCOs of charged particles around a Kerr black hole with a weak magnetic field, originally discovered by Wald in \cite{wald}. In contrast to previous research, we focus on the case $\mathcal{B}=\frac{qB}{2m}\ll1$, where $q$, $m$, and $B$ are the charge, mass, and strength of the magnetic field. This condition allows us to obtain an analytical expression for the radius of the ISCO that involves the energy $E$ and angular momentum $L$ of charged particles. As many astrophysical clouds of dust or fluids in our universe do not have an extreme charge-mass ratio, the ubiquitous satisfaction of the  $\mathcal{B}\ll1$ condition makes our results useful in practical situations. In addition, we examine the polarized image of charged particles in circular motion on the equatorial plane observed by an observer at infinity. Unlike typical treatments of synchrotron radiations of a charged particle around a black hole, which rely on radiation study in the local inertial frame, we apply the covariant formulas of electromagnetic radiations that encode the intensity and polarization direction of radiations in curved spacetime. These formulas were recently developed in \cite{Hu:2022sej}, building on earlier related discussions found in \cite{Quinn:1996am}. In addition, it is relevant to note that we have developed a simplified model in which charged particles move in circular orbits on the equatorial plane under the combined influence of gravity and Lorentz force, reminiscent of electrons in a ring accelerator. Such motion of charged particles in a magnetic field produces linearly polarized radiation, which becomes synchrotron radiation when the particles move relativistically \cite{Hu:2022sej}. However, when observing a black hole in a real astrophysical system, the EHT has observed synchrotron radiation originating from magnetic accretion flows \cite{EventHorizonTelescope:2021bee, EventHorizonTelescope:2021srq, narayan2021polarized, Himwich:2020msm}. Magnetic fluid is a plasma made up of charged particles and possesses a magnetic field. As these charged particles spiral within the magnetic field at relativistic speeds, they emit synchrotron radiation. Because of this difference, our simplified model cannot entirely account for the polarized images of black holes observed so far, and thus is more suited for theoretical investigation. Nevertheless, if polarized radiation from charged particles can be directly observed, our model will become more applicable. Building on our simplified model, we demonstrate the polarization directions and total flux of the radiation emitted from charged particles for various spins, observational angles, and orbits. Our numerical studies on polarized images are not restricted to the case $\mathcal{B}\ll1$, although our primary focus is on researching innermost stable circular orbits (ISCOs) within this approximation.

The remaining parts of this paper are organized as follows. In Sec. \ref{sec2}, we study the circular orbits of the charged particles around a Kerr black hole with a Wald magnetic field. In Sec. \ref{sec3}, we review the electromagnetic radiations in curved spacetime\cite{Hu:2022sej} and calculate the polarized images of the charged particles in circular motions around a Kerr black hole. The main conclusions are summarized in Sec. \ref{summary}. In this work, we have set the fundamental constants $c$, $G$ and the vacuum permittivity $\varepsilon_0$ to unity, and we will work in the signature convention $(-, +, +, +)$ for the spacetime metric.

\section{Circular orbits of a charged particle around a Kerr black hole with a Wald magnetic field}\label{sec2}
In this section, we would like to focus on the circular orbits of a charged particle around a Kerr black hole with a Wald magnetic field.

\subsection{Kerr black hole with a Wald magnetic field}
The spacetime of a neutral and rotating black hole in general relativity is described by the Kerr black hole metric, which is a stationary and axisymmetric solution of the Einstein equations and takes,
\bea
\mathrm{d}s^2=-\left(1-\frac{2Mr}{\Sigma}\right)\mathrm{d}t^2+\frac{\Sigma}{\Delta}\mathrm{d}r^2+\Sigma\mathrm{d}\theta^2+\left(r^2+a^2+\frac{2Mra^2}{\Sigma}\mathrm{sin}^2\theta\right)\mathrm{d}\phi^2-\frac{4Mra}{\Sigma}\mathrm{sin}^2\theta\mathrm{d}t\mathrm{d}\phi\,,\nn\\
\eea
in the Boyer-Lindquist coordinates, where
\bea
\Sigma=r^2+a^2\mathrm{cos}^2\theta,\quad \Delta = r^2-2Mr+a^2\,,
\eea
with $M$ and $a$ being the mass and spin of the Kerr black hole, respectively. And the angular momentum of the Kerr black hole is $J=Ma$. The horizons are located at $r_\pm=M\pm\sqrt{M^2-a^2}$, given by the equation $\Delta=0$. The outer one is the event horizon of the Kerr black hole, that is, $r_h=r_+$. Next, we would like to assume that there is a vertical and uniform magnetic field outside the Kerr black hole, described by the Wald solution to the source-free Maxwell equations, that is, $\nabla_\mu F^{\mu\nu}=0$. The magnetic field is aligned with the rotation axis of the black hole and independent of the coordinates $t$ and $\phi$. Considering the two Killing vectors $\eta^a=\left(\frac{\partial}{\partial t}\right)^a$ and $\psi^a=\left(\frac{\partial}{\partial \phi}\right)^a$, Wald proved that the field
\be
A^a=\frac{1}{2}B(\psi^a+\frac{2J}{M}\eta^a)-\frac{Q}{2M}\eta^a \label{A}
\ee
gives a unique field that is an asymptotically uniform magnetic field of strength $B_0$, with $Q$ being the charge of the Kerr black hole. Next, we introduce the electrostatic injection energy, defined as the difference between the electric potential at the horizon and infinity, i.e.,
\bea
\epsilon=e A_\mu\eta^\mu|_{\text{horizon}}-eA_\mu\eta^\mu|_\infty
\eea
which is a constant over the black hole\cite{wald}. For a pure Kerr spacetime, we should have $Q=0$. However, due to non-vanishing electrostatic injection energy $\epsilon$, an electric field is induced by the rotation of the black hole in the presence of the magnetic field. It causes charged particles to be attracted by the black hole, a process called charged accretion. As a result, to reach a steady state, the black hole has to gain a nonzero net charge to reach equilibrium, that is, $\epsilon=0$, which gives
\begin{equation}\label{ec}
Q=2BJ\,.
\end{equation}
In the following, we would like to focus mainly on this equilibrium situation. Note that in \cite{al2022charged}, they also paid attention to the non-equilibrium situations. It is worth mentioning that Eq. (\ref{ec}) holds for an arbitrary stationary, axisymmetric black hole; that is, the condition is not limited to Kerr black holes. In principle, the charge and magnetic field would react upon the background spacetime. If we want to ignore the contribution of the magnetic field to the background spacetime for simplicity, we need to consider the weak field approximation,
\bea
BM\ll 1\,.
\eea
On the other hand, the characteristic length scale given by the charge of the black hole is comparable with the gravitational radius when
\bea\label{qg}
\sqrt{\frac{Q_G^2G}{c^4}}=\frac{2GM}{c^2}\,,
\eea
so that the gravitational effect of the charge $Q$ on the background can be ignored if
\bea
Q\ll Q_G\,.
\eea
Note that we have restored $G$ and $c$ in Eq. (\ref{qg}), making the physical meaning of the formula more explicit.
In addition, considering $J\le M^2$, from Eq. (\ref{ec}) we have
\bea
\frac{Q}{Q_G}=\frac{Q}{2M}=Ba\le BM\ll1\,,
\eea
which means we are allowed to omit the charge's influence on the spacetime metric, although there is a nonzero net charge so that we can stick to the Kerr metric under the weak field approximation. In fact, in the Gauss units, we find
\bea\label{btg}
B_{\text{Gauss}}=\frac{c^4}{G^{3/2}M}(BM)\simeq2.36\times 10^{19}\frac{M_{\odot}}{M}(BM)\quad \text{Gauss}\,.
\eea
Here (BM) is a dimensionless number and can also be seen as the value of $B$ in our simplified unit after setting $c$, $G$, $\epsilon_0$ and $M$ to 1, and $M_\odot$ is the mass of the sun. For supermassive black holes $M\sim 10^9 M_\odot$, the dimensionless $BM=0.01\ll 1$ corresponds to $B_{\text{Gauss}}\sim 10^8 \text{Gauss}$, which is already a very significant value for astrophysically appropriate magnetic field around an astronomical black hole. Thus, the weak field approximation $BM\ll 1$ can be imposed for most real astronomical black holes with magnetic fields and deserves careful study.

\subsection{Motions of charged particles on the equatorial plane}
In this subsection, we move to study the motions of charged particles on the equatorial plane under the weak field approximation employing the standard approach. In the Boyer-Lindquist coordinates, the generalized four-momentum $P^\mu$ can be written as
\begin{equation}
P^\mu=p^\mu+qA^\mu\,,
\end{equation}
where $p^\mu=mu^\mu$ is the four-momentum for the charged particle with $m$ being the mass, $A^\mu$ is the electric vector potential and $q$ is the charge of the particle. From \eqref{A} we have
\begin{equation}
A^\mu=(aB-\frac{Q}{2M},0,0,\frac{B}{2})\,,
\end{equation}
Note that the Lie derivatives of $A_\mu$ with respect to the Killing vectors $\eta^\mu$ and $\psi^\mu$ are zero, that is,
\bea
\mathcal{L}_\eta A_\mu=0\,,\quad \mathcal{L}_\psi A_\mu=0\,,
\eea
thus, we can construct two conserved quantities along the motions of charged particles
\bea\label{ccs}
E&=&-\frac{P^\mu}{m}g_{\mu\nu}\eta^\nu=-P_t/m\,,\nn\\
L&=&\frac{P^\mu}{m}g_{\mu\nu}\psi^\nu=P_\phi/m\,.
\eea
Note that the energy can be explicitly written out as
\begin{gather}
E=\left(1-\frac{2Mr}{\Sigma}\right)\left[\frac{q\left(aB-\frac{Q}{2M}\right)}{m}+\frac{\mathrm{d}t}{\mathrm{d}\tau}\right]+\frac{aMr\mathrm{sin}^2\theta\left(Bq+2m\frac{\mathrm{d}\phi}{\mathrm{d}\tau}\right)}{m\Sigma}\,,
\end{gather}
so it is not necessarily positive. Similarly, $L$ wouldn't be necessarily positive even if $\frac{\partial \phi}{\partial \tau}>0$.

\begin{figure}
\centering
\includegraphics[width=8cm]{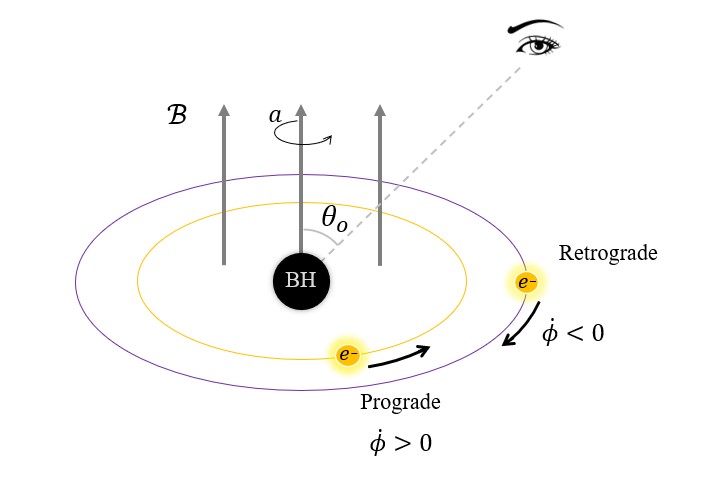}
\caption{A diagram of charged particles moving in circular orbits on the equatorial plane of the Kerr spacetime with a vertical and uniform magnetic field. We assume the spin of the Kerr black hole and the magnetic field are in the same direction when $a$ and $\mathcal{B}$ have the same sign.}\label{setup}
\end{figure}

Since we are interested in the circular motions confined to the equatorial plane as illustrated in Fig. \ref{setup}, the four-velocity of the charged particle takes
\bea
u^\mu=(\dot{t}, \dot{r}, 0, \dot{\phi})\,,
\eea
where `` $\cdot{}$ '' denotes derivative with respect to the proper time $\tau$. From the Eq. (\ref{ccs}), we can obtain
\begin{gather}
\frac{\mathrm{d}t}{\mathrm{d}\tau}=\frac{g_{\phi\phi}(E+\frac{q}{m}A_t)+g_{t\phi}(L-\frac{q}{m}A_\phi)}{g^2_{t\phi}-g_{tt}g_{\phi\phi}}\,, \\
\frac{\mathrm{d}\phi}{\mathrm{d}\tau}=\frac{g_{tt}(-L+\frac{q}{m}A_t)+g_{t\phi}(-E-\frac{q}{m}A_t)}{g_{t\phi}^2-g_{tt}g_{\phi\phi}}
\end{gather}
Note that the four-velocity of a charged particle is future-directed, thus we have $\dot{t}>0$ which gives us
\bea\label{lle}
E>\frac{2aL}{r^3+a^2r+2a^2}\,,
\eea
where we have used the equilibrium condition, that is, $Q=2BJ$, and set the mass of the black hole $M=1$ here and hereafter. Moreover, the four-velocity of the charged particle is normalized, that is, $u^\mu u_\mu=-1$. Then, combining the Eq. (\ref{ccs}) and the normalization condition of the four-velocity, we can obtain
\bea
\dot{r}^2+V_{eff}=0\,,
\eea
where we have defined the effective potential $V_{eff}$ as
\bea\label{pre}
V_{eff}=\frac{\mathcal{B}^2\Delta(a^2r+r^3+2a^2)-2Lr\Delta\mathcal{B}-(2a^2+a^2r+r^3)E^2+4aLE +r\Delta -2L^2+L^2r}{r^3}\,.\nn\\
\eea
Here, we have introduced a new parameter
\bea
\mathcal{B}=\frac{qB}{2m}\,,
\eea
It can be explicitly written as $V_{eff}=\frac{P_5(r)}{r^3}$, where $P_5(r)$ is a fifth-order polynomial of $r$. To write out the coefficients explicitly, we have
\begin{gather}
P_5(r)=c_5 r^5+c_4 r^4+c_3 r^3+c_2 r^2+c_1 r +c_0\,,\nn\\ \label{p5}
c_0=-2(L-aE)^2+2a^4 \mathcal{B}^2 \,,\nn \\
c_1=a^2+(-4a^2+a^4)\mathcal{B}^2-2a^2\mathcal{B}L+L^2-a^2E^2 \,,\nn\\
c_2=-2+4\mathcal{B}L \, ,\,
c_3=1+2a^2\mathcal{B}^2-2\mathcal{B}L-E^2\,,\nn\\
c_4=-2\mathcal{B}^2 \,,\,
c_5=2\mathcal{B}^2
\end{gather}

In addition, it will useful to evaluate $V_{eff}$ at the horizon $x=\frac{1}{r_h}$ and the result reads
\bea
-r_+^3V_{eff}(r=r_h)=4E^2r_++L^2r_--4aEL
\eea
Here the terms regarding $\mathcal{B}$ have become zero. This also implies that around the horizon, the impact of the magnetic field on particles would become small. Also,
\begin{equation}
4E^2r_++L^2r_--4aEL\ge2\sqrt{4E^2L^2r_+r_-}-4aEL=0\,,
\end{equation}
therefore, we can see that the effective potential is always non-negative at the horizon. The equation holds if and only if $4E^2r_+=L^2r_-$, which gives $\frac{L}{E}=\pm\frac{2r_+}{a}$.

\subsection{Circular orbits of charged particles}
Next, we turn to calculate the circular orbits of charged particles in the Kerr spacetime with a Wald magnetic field. For simplicity, we would like to set $\mathcal{B}>0$ and let the spin $a$ vary from $-1$ to $1$. To find the energy $E$ and angular momentum $L$ at a radius $r$, we need to solve the following equations
\bea\label{coe}
V_{eff}=0\,\quad\text{and} \quad\,\partial_rV_{eff}=0\,,
\eea
and the ISCO can be determined by further requiring
\bea\label{iscoe}
\partial_r^2 V_{eff}=0\,.
\eea
It's easy to see that $V_{eff}=0$ can be reduced to a quintic equation for $r$, which cannot generally be solved analytically. And thus, one has to solve Eqs (\ref{coe}) and (\ref{iscoe}) numerically to gain insights into the ISCOs of charged particles in the Kerr spacetime with a Wald magnetic field, see examples in \cite{al2022charged}.

However, one can see that $V_{eff}=0$ reduces to a cubic equation for $r$ if the term including $\mathcal{B}^2$ can be dropped, which is legal when $\mathcal{B}\ll1$. Considering that a cubic equation can be solved analytically and an analytical expression is useful for better understanding the circular orbits of charged particles in the Kerr spacetime with a Wald magnetic field, we would like to carefully study the case $\mathcal{B}\ll1$. Let us start by discussing when the approximate condition holds in the Gauss units. Note that in the present convention, $\mathcal{B}=k$ means $\mathcal{B}=kM^{-1}$, where $k$ is a dimensionless number, we can recover the full expression of $\mathcal{B}=k$ in the Gauss units,
\bea
\left(\frac{qB}{2m}\right)_{\text{Gauss}}= k\frac{c^4}{GM}\,,
\eea
and this gives
\bea
\left(\frac{q}{e}\right)\left(\frac{m_p}{m}\right)=\frac{43}{B_{\text{Gauss}}}\frac{M_\odot}{M}k=1.8\times10^{-18}\frac{(\mathcal{B}M)}{(BM)}\,,
\eea
where $m_p$ is the proton mass, $e$ is the unit charge and we have plunged Eq. (\ref{btg}) in the last `` $=$ ''. Thus, the approximate condition $\mathcal{B}\ll1$ corresponds to
\bea\label{ca}
\left(\frac{q}{e}\right)\left(\frac{m_p}{m}\right)(BM)\ll 1.8\times 10^{-18}\,.
\eea
As a result, we can see that the approximate condition $\mathcal{B}\ll1$ is applicable for an object with a unit charge whose mass is much larger than $10^{18}(BM)m_p$. For example, from Eq. (\ref{btg}), we find $BM=10^{-9}$ if we consider a supermassive black hole with $M\sim10^9M_{\odot}$ immersed in a magnetic field with $B=10\, \text{Gauss}$, then the mass of the object with a unit charge should be much larger than $m=10^{9}m_p=10^{-18}\,\text{kg}$ in the International System of Units if we require that the approximate condition $\mathcal{B}\ll1$ holds in our present convention. To be precise, the charge-to-mass ratio of an electron is $1.7\times10^{11}$, which is too extreme for our approximation in Eq. (\ref{ca}), as an electron would have $\mathcal{B}=8.6\times10^{10}$. However, the charge-to-mass ratio of ionized gas molecules is much lower, ranging from $10^6$  to $10^8$, making it easier to satisfy the approximation condition $\mathcal{B}\ll1$. This approximation is particularly helpful for studying charged astrophysical dust clouds and weakly magnetized black holes with low mass.

Now, we move to study the orbits of charged particles in the case $\mathcal{B}\ll1$. We dropped the second order terms for $\mathcal{B}$ in the effective potential, and the Eq. (\ref{pre}) can be simplified as
\bea
-V_{eff}=2(L-2aE)^2x^3+(-a^2+2a^2\mathcal{B}L-L^2+a^2E^2)x^2+(2-4\mathcal{B}L)x+E^2+2\mathcal{B}L-1\,,\nn\\
\eea
where a new parameter $x=1/r$ is introduced for simplicity. The equations for a circular orbit
\bea
V_{eff}=\partial_r V_{eff}=\partial_r^2V_{eff}=0
\eea
correspond to the equations
\bea\label{epx}
V_{eff}=\partial_x V_{eff}=\partial_x^2V_{eff}=0\,,
\eea
This would be true except for two points: $r=0$ inside the horizon and $r=\infty$, so we can obtain the ISCOs by solving the Eq. (\ref{epx}).

In the following subsections, we will analyze the effective potential's form and root structures under this approximation.

\subsubsection{Quadratic form}
Next, let us begin with the first equation, $V_{eff}=0$. It's easy to see that the cubic equation is simplified into a quadratic equation when $L=aE$. Then the effective potential function becomes \
\begin{align}
-V_{eff}&=(-1+2\mathcal{B}L)a^2x^2+2(1-2\mathcal{B}L)x+E^2+2\mathcal{B}L-1 \\
&=(-1+2\mathcal{B}L)\left(ax-\frac{1}{a}\right)^2+E^2+(2\mathcal{B}L-1)\left(1+\frac{1}{a^2}\right) \, .
\end{align}
When $-1+2\mathcal{B}L=0$, that is, $L=\frac{1}{2\mathcal{B}}$, we find $V_{eff}=-E^2$, which implies no stable circular orbits of charged particles exist in this situation.

When $-1+2\mathcal{B}L>0$, we find
\bea
-V_{eff}(x=0)=E^2+2\mathcal{B}L-1>0\,, -V_{eff}(x=\frac{1}{r_+})>0\,,\frac{1}{a^2}\ge1\ge\frac{1}{r_+}\,,
\eea
so that $-V_{eff}$ is a monotonic decreasing function and $-V_{eff}>0$ is always true in the region $x\in(0, 1/r_+)$, which implies there are no circular orbits of charged particles.

At last, we focus on the case $2\mathcal{B}L-1<0$. In addition, combining the Eq. (\ref{lle}) with $L=aE$, we can find the energy satisfies
\bea
0<E<\frac{1}{2\mathcal{B}a}\quad\text{with}\quad 0<a\le1\,,
\eea
thus, we can obtain
\bea
-1<-V_{eff}(x=0)=E^2+2\mathcal{B}aE-1<\frac{1}{4\mathcal{B}^2a^2}\,.
\eea
On the other hand, we find $-V_{eff}$ is a monotonic increasing function in the region $x\in(0, 1/r_+)$ in the case $2\mathcal{B}L-1<0$, and so
\bea
-V_{eff}(x=0)<-V_{eff}<-V_{eff}(x=1/r_+)
\eea
in the region $x\in(0, 1/r_+)$. Thus, when $0<E<\sqrt{1+a^2\mathcal{B}^2}-a^2\mathcal{B}^2 $ we have $-V_{eff}(x=0)<0$ and there must be a $r_0$, such that $V_{eff}(r_0)=0$. This implies there's a forbidden zone $(r_0, +\infty)$ for charged particles when $0<E<\sqrt{1+a^2\mathcal{B}^2}-a^2\mathcal{B}^2$. When $\sqrt{1+a^2\mathcal{B}^2}-a^2\mathcal{B}^2\le E<\frac{1}{2\mathcal{B}a}$, we have $-V_{eff}(x=0)>0$ and there's no forbidden zone for charged particles. In both cases, there are no circular orbits for charged particles.

In conclusion, now we can ensure that there are no circular orbits for charged particles when $L=aE$.

\subsubsection{Cubic form}
Then let's pay attention to the more general situation in which $L\neq aE$ and the effective potential function is cubic. To investigate the structure of the potential, it's convenient to define a new variable $y$ by
\bea
x=y+\frac{d}{3\tilde{a}}\,,
\eea
where we set
\bea
d&=&L^2+a^2(1-2\mathcal{B}L-E^2)\,,\nn\\
\tilde{a} &=& 2(L-aE)^2\,,
\eea
and then we get a compact form
\bea
f(y)=-V_{eff}=\tilde{a}y^3+by+c\,,
\eea
where
\bea
b &=& 2-4\mathcal{B}L-\frac{d^2}{3\tilde{a}}\,,\nn\\
c &=& -1+2\mathcal{B}L+E^2+\frac{2(1-2\mathcal{B}L)d}{3\tilde{a}}+\frac{d^3}{27\tilde{a}}\,.
\eea
Then, we find
\bea
-\partial_xV_{eff}(x)=\partial_yf(y)=3\tilde{a}t^2+b\,.
\eea
A circular orbit has to satisfy $\partial_xV_{eff}(x)=0$, such that we have $b<0$ since $\tilde{a}$ is always positive. Furthermore, $y=\sqrt{\frac{-b}{3\tilde{a}}}$ gives us a local maximum of the effective potential $V_{eff}$ while $y=-\sqrt{\frac{-b}{3\tilde{a}}}$ corresponds to a local maximum. Then the necessary and sufficient conditions for the existence of a future-directed stable circular orbit are
\begin{equation}\label{cd2}
0<y_0=\sqrt{\frac{-b}{3\tilde{a}}}+\frac{d}{3\tilde{a}}<\frac{1}{r_+}\,,
\end{equation}
and
\begin{equation} \label{ss}
f(\sqrt{\frac{-b}{3\tilde{a}}})=0\,,
\end{equation}
as well as the Eq. (\ref{lle}). In addition, the condition $f(\sqrt{\frac{-b}{3\tilde{a}}})=0$ is equivalent to
\bea\label{cd1}
\frac{c^2}{4}+\frac{b^3}{27\tilde{a}}=0\,,
\eea
which gives a constraint for the conserved quantities $L$ and $E$. That is to say when the Eqs. (\ref{cd2}), (\ref{cd1}) and (\ref{lle}) hold, a stable future-directed circular orbit for a charged particle must exist, and the radius is given by
\bea
r_0=1/y_0\,,
\eea
where $y_0$ is given in Eq. (\ref{cd2}).

We can push our calculations to determine the ISCO of the charged particle. Considering the equation $\partial_x^2V_{eff}(x)=\partial_y^2f(y)=0$, we can easily find the following conditions
\bea\label{bcz}
b&=&c=0\,,\\
r_{ISCO}&=&1/y_{ISCO} =\frac{3\tilde{a}}{d}=\frac{6(L-aE)^2}{a^2-2a^2\mathcal{B}L+L^2-a^2E^2}\,,\label{iscor}
\eea
are true if there's an ISCO of the charged particle on the equatorial plane in the Kerr spacetime with a weak magnetic field under the approximate condition $\mathcal{B}\ll1$. Here, it is worth emphasizing that the energy $E$ and angular momentum $L$ can be obtained by solving the Eq. (\ref{bcz}), and then the Eq. (\ref{iscor}) can give us the value of the radius of the ISCO. While, for $\mathcal{B}\gg1$ and $\mathcal{B}\sim\mathcal{O}(1)$, we have to obtain the ISCO and orbits of charged particles numerically. Note that in numerical calculations, we can use the fact that if Eqs. (\ref{coe}) and (\ref{iscoe}) hold, then the fifth order polynomial \ref{p5} that we mentioned should also satisfy
\bea
P_5(r)=0\, , \, \partial_rP_5(r)=0
\eea
which will greatly simplify the computation.

\section{Polarized images of charged particles in circular orbits}
\label{sec3}
In this section, we move to study polarized images of charged particles that move in circular motions on the equatorial plane of the Kerr spacetime with a Wald magnetic field. The observer is located at infinity with an inclination angle $\theta_o$ as shown in Fig. \ref{setup}.

\subsection{Polarization of electromagnetic radiations and propagation of light}
In this subsection, we would like to introduce the necessary formulas to calculate the polarization of electromagnetic radiations originating from the charged particle and propagation of light following our previous work \cite{Hu:2022sej}. When charged particles are affected by Lorentz forces, they are obliged to move in non-geodesic trajectories and generate electromagnetic radiations with nice polarization properties. The polarization vector of the electromagnetic radiation takes
\bea\label{pve}
f^\mu=N^{-1}\left[\delta^\mu_\alpha\left(k_\beta D_\tau u^{[\beta}u^{\alpha]}\right)\right]\,,
\eea
where $u^\alpha$ has been introduced to be the four-velocity of the charged particle in the last section, $k_\beta$ is the wave vector of the polarized radiation and normalized by $k^\beta u_\beta=1$, $D_\tau$ is derivative operator along the vector $u^\beta$, that is, $D_\tau\equiv u^\alpha\nabla_\alpha$ with $\tau$ being the proper time of $u^\alpha$, and $N^{-1}$ is a normalization factor in ensuring $f_\mu f^\mu=1$. In addition, the luminosity of the polarized radiation is given by
\bea
L=4\pi|k_\beta D_\tau u^\beta u^\alpha-D_\tau u^\alpha|^2\,.
\eea
Next, we consider the photon's position that hits the observer's screen with coordinates $(t_o, r_o, \theta_o, \phi_o)$. In terms of celestial coordinates, the position can be described by
\bea
\alpha=-\frac{\lambda}{\sin\theta_o}\,,\quad\beta=\pm_o\sqrt{\Theta(\theta_o)}\,,
\eea
where $\pm_o=\text{sign}(k_o^\theta)$ represents the sign of $k^\theta$ in the observer's frame, $\lambda=l/\omega$ is an impact parameter with $l$ and $\omega$ being the angular momentum and energy of the photon, respectively. $\Theta(\theta)$ is the angular potential of the photon, which reads
\bea
\Theta(\theta)=\eta+a^2\cos^2\theta-\lambda^2\cot^2\theta\,,
\eea
where $\eta=\mathcal{Q}/\omega^2$ is the other impact parameter with $\mathcal{Q}$ being the Carter constant in the Kerr spacetime. On the other hand, charged particles as the light source are located on the equatorial plane with coordinates $(t_s, r_s, \pi/2, \phi_s)$. In principle, with the help of null geodesic equations in Kerr spacetime, one can obtain the trajectories of photons connecting the source and the observer, and then we can find the images of the source; that is, we know the coordinates $(\alpha, \beta)$ on the screen of the observer. We want to stress that there is more than one trajectory of lights starting at the source and ending at the observer since the lights may turn around the black hole more than once before reaching the observer. We use $m$ to denote the number of times the trajectory crosses the equatorial plane between the source and the observer; different values of $m$ correspond to the $(m+1)$th image on the observers' screen. In particular, the image formed by the trajectories with $m=0$ is often called the primary image, the image with $m=1$ is called the secondary image, and so on. For more details, one is suggested to refer to \cite{Hou:2022gge}.

Furthermore, we also need to obtain the polarization information of the image. Following \cite{Hu:2022sej}; the polarization information can be transmitted from the source to the observer by using the Penrose-Walker (PW) constant
\bea\label{kappa}
\kappa=2k^\mu f^{\nu}\left(\hat{l}_{[\mu}\hat{n}_{\nu]}-\hat{m}_{[\mu}\hat{\bar{m}}_{\nu]}\right)(r-ia\cos\theta)\equiv\omega(\kappa_1+i\kappa_2)\,,
\eea
which is a conserved quantity along the null geodesics for Kerr spacetime. In the above formula, $\kappa_{1,2}$ are the rescaled real and imaginary parts of the PW constant, respectively. And ${\hat{l}, \hat{n}, \hat{m}, \hat{\bar{m}}}$ are the Neumann-Penrose tetrads which can be chosen as
\bea
\hat{l}&=&\frac{1}{\sqrt{2\Delta}\Sigma}\left[(r^2+a^2)\partial_t+\Delta\partial_r+a\partial_\phi\right]\,\nn\\
\hat{n}&=&\frac{1}{\sqrt{2\Delta}\Sigma}\left[(r^2+a^2)\partial_t-\Delta\partial_r+a\partial_\phi\right]\,\nn\\
\hat{m}&=&\frac{1}{\sqrt{2}(r+ia\cos\theta)}\left(ia\sin\theta\partial_t+\partial_\theta+\frac{i}{\sin\theta}\partial_\phi\right)\,,\nn\\
\hat{\bar{m}}&=&\frac{1}{\sqrt{2}(r-ia\cos\theta)}\left(-ia\sin\theta\partial_t+\partial_\theta+\frac{i}{\sin\theta}\partial_\phi\right)\,.
\eea
Thus, employing the Eq. (\ref{kappa}), at the observer, we can decode the information of $f^\mu$ given at the source and on the polarization, the vector can be read from the PW constant as
\begin{equation}
\overrightarrow{\mathcal{E}}=\left(\mathcal{E}_\alpha, \mathcal{E}_\beta\right)=\frac{1}{\omega\left(\beta^2+\gamma^2\right)}\left(\beta \kappa_2-\gamma \kappa_1, \beta \kappa_1+\gamma \kappa_2\right), \quad \gamma=-\left(\alpha+a \sin \theta_o\right),
\end{equation}

On the other hand, we still need to know the total fluxes of the image considering charged particles have a finite size. We omit the details and give the result of the total fluxes instead, which takes
\bea
F_o=\frac{g^4L}{4\pi r_o^2|\hat{k}|\cdot X}\left|\frac{\partial(\alpha, \beta)}{\partial(Y_s, Z_s)}\right|\,,
\eea
where $\left|\frac{\partial(\alpha, \beta)}{\partial(Y_s, Z_s)}\right|$ denotes the Jacobian determinant between the coordinates $(\alpha, \beta)$ and $(Y_s, Z_s)$. Thereinto, $(Y_s, Z_s)$ are coordinates defined on the `` source screen '' with $T=X=0$. $\{T, X, Y, Z\}$ is a local Minkowski coordinate system in the neighborhood of the source, which is defined as
\bea
Te_{(t)}+Xe_{(r)}+Ye_{(\phi)}-Ze_{(\theta)}=(x^\mu-x_\ast^\mu)\partial_\mu\,,
\eea
where $x_\ast^\mu$ are the coordinates of the source and $e_{(\mu)}$ are the tetrad of the source. For a Kerr metric, we would like to choose
\bea
e_{(t)}&=&\gamma\sqrt{\frac{\Xi}{\Delta\Sigma}}(\partial_t+\Omega_s\partial_\phi)\,,\quad e_{(r)}=\frac{\Delta}{\Sigma}\partial_r\,,\quad e_{(\theta)}=\frac{1}{\sqrt{\Sigma}}\partial_\theta\,,\nn\\
e_{(\phi)}&=&\gamma v_s\sqrt{\frac{\Xi}{\Delta\Sigma}}\left[\partial_t+\frac{2aMr}{\Xi}\right]+\gamma\sqrt{\frac{\Sigma}{\Xi}}
\eea
where we have introduced
\bea
\Xi=(r^2+a^2)^2-\Delta a^2\sin^2\theta\,,
\eea
and
\bea
v_s=\frac{\Xi}{\Sigma\sqrt{\Delta}}(\Omega_s-\frac{2aMr}{\Xi})\,,\quad\gamma=\frac{1}{\sqrt{1-v_s^2}}\,,
\eea
with $\Omega_s=\frac{d\phi_s}{dt_s}$ being the angular velocity of the source. In addition, the unit vector $\hat{k}$ is given by
\bea
\hat{k}=\frac{1}{k^{(t)}}\left(k^{(r)}X+k^{(\phi)}Y-k^{(\theta)}Z\right)\,,
\eea
and $g$ is the redshift factor which is given by
\bea
g=\frac{\omega_o}{\omega_s}\,,
\eea
where $\omega_{o,s}$ are the frequencies of the photon in the observer and the source frame, respectively. Now, we are ready to calculate some specific polarized images of charged particles orbiting the Kerr black hole with a Wald magnetic field.

\subsection{Results}
Based on our model, in this subsection, we would like to show some results for our calculations with the observation angle at $\theta_o=17^\circ$ and $\theta_o=80^\circ$. In our computations of the polarized images, we use the full form of the effective potential to numerically determine the parameters for stable circular orbits without approximation. According to the previous analysis, when $\mathcal{B}$ and $a$ are given, only one degree of freedom is left among $r$, $E$, and $L$ for stable circular orbits. In our following calculations, we usually set $r_s=6$ and compare the results with other different parameters; that is, we fix the radius of the orbit for charged particles and discuss the influence of the other parameters. And the reason why we choose $r_s=6$ is that the ISCO for a neutral particle in a Schwarzschild black hole spacetime is at $r_s=6$, and we still hope to focus on polarized images of the source in this region. Our main interest is in the effects of the spin $a$, the observational angle $\theta_o$ on the polarized images, and the difference between the
retrograde or prograde orbiting source \footnote{Note that prograde (retrograde) is defined by $\dot{\phi}>0$ ($\dot{\phi}<0$) instead of $L>0$ ($L<0$), since the sign of the angular momentum $L$ is not necessarily as same as the sign of $\dot{\phi}$, when there is an external magnetic field.}.
In the following studies, we will vary $a$ and $\theta_o$ and make plots in retrograde or prograde cases. In addition, as pointed out in \cite{Hu:2022sej}, empirically, the strength of $\mathcal{B}$ doesn't affect the direction of the polarization vector $f^\mu$ and only changes the overall intensity of flux. It's also the case with Kerr spacetime. So we also fix $\mathcal{B}=2$ in our work.

\begin{figure}[h!]
\centering
\includegraphics[width=15cm]{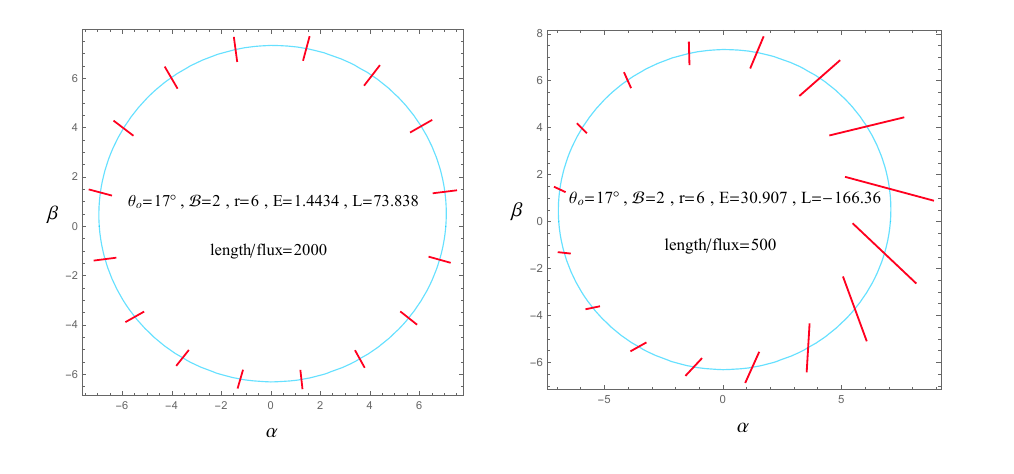}
\caption{Comparisons of polarized images between the prograde and retrograde orbits with $\theta_o=17^\circ$, $a=0.94$, $r_s=6$ and $\mathcal{B}=2$. The left plot is for prograde orbit, while the right is retrograde. The corresponding energy $E$ and angular momentum $L$ for stable circular orbits are also shown in each plot. Blue circles give shapes of the images of charged particles, and red lines denote polarization information. In particular, the length of the red line signifies the magnitude of the total flux. Note that the lengths of the flux are scaled differently to make the lengths of lines in the two plots of the same order of magnitude.}\label{figure:rp}
\end{figure}

At first, we look at the difference in the polarized images between the prograde and retrograde orbits. We would like to fix $a=0.94$, which is a potential value of the spin of the supermassive black hole in M87 predicted by EHT \cite{EventHorizonTelescope:2019dse, EventHorizonTelescope:2019jan, EventHorizonTelescope:2019ths, EventHorizonTelescope:2019pgp, EventHorizonTelescope:2019ggy}. The results are shown in Fig. \ref{figure:rp}, in which blue circles are the images of charged particles and red lines denote the polarization directions and total flux with the length. From Fig. \ref{figure:rp}, one can see that for the prograde case in the left panel, there is barely any difference between the lengths of red lines on the left and right side. However, for the retrograde case in the right forum of Fig. \ref{figure:rp}, we can find that the fluxes are much stronger on the right side than those on the left side. The main reason for this phenomenon is the so-called Doppler beaming (sometimes called the headlight effect). The radiations from moving particles concentrate upon the vicinity along the direction of the motion of particles, which becomes more significant when the velocity of moving particles is higher. Qualitatively, when $\mathcal{B}$ and $L$ are in opposite directions, charged particles would have higher energies to stay in a circular orbit so that the Doppler beaming is more substantial in the retrograde case. In addition, we can infer that the phenomenon would be more distinct when the observer is closer to the equatorial plane.

\begin{figure}[h!]
\centering
\includegraphics[width=15cm]{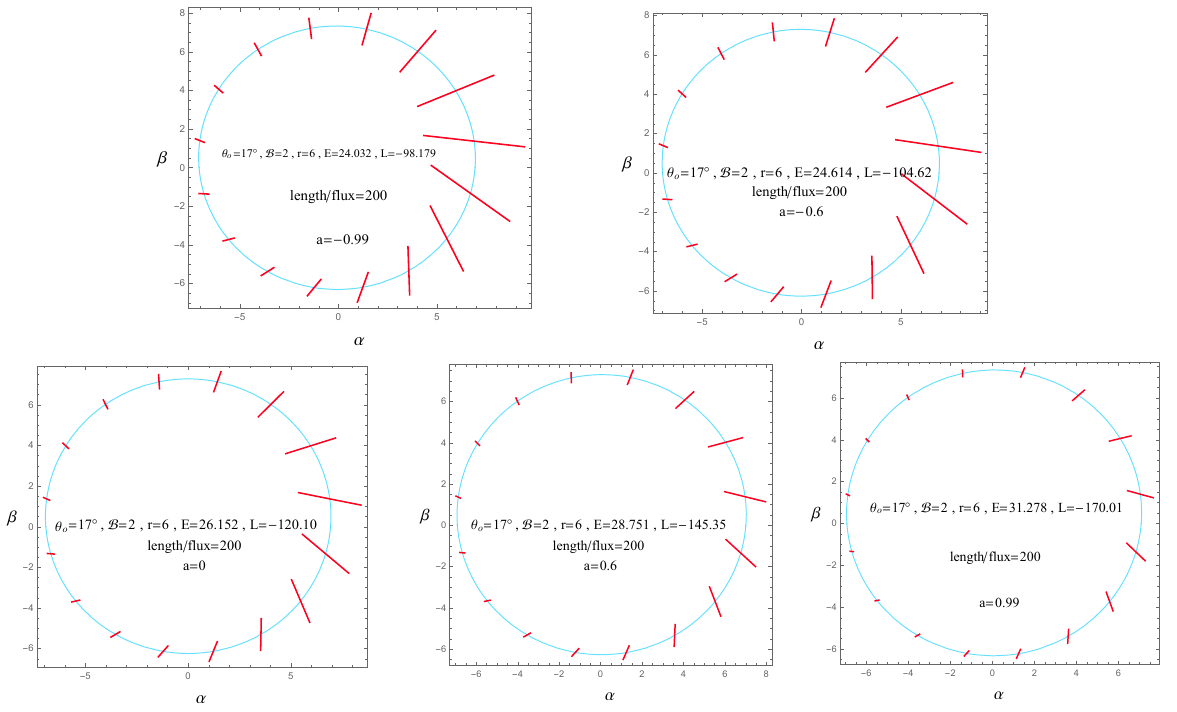}
\caption{Polarized images of charged particles in retrograde orbits are produced with $\theta_o=17^\circ$, $\mathcal{B}=2$, and $r_s=6$. We let the spin $a$ vary from $-0.99$ to $0.99$. The energy $E$ and the angular momentum $L$ are directly calculated from the conditions of the stable circular orbit. Blue lines show the shapes of the images of the stable orbits. The inclinations of red lines denote the polarization directions of radiations, while the lengths represent the total flux. In this set of plots, we set length/flux = 4000 to make the plots more ornamental.}\label{figure:a}
\end{figure}

Next, we turn to study the influence of the spin on the polarized images. In Fig. \ref{figure:a}, we fix $\theta_o=17^\circ$, $r_s=6$ and vary $a$ for retrograde orbits. The length ratio to the flux is $200$ for the five plots. We can see evident changes in the magnitude of the flux, while the corresponding polarized directions show no significant changes as the spin $a$ increases. In addition, we also checked the prograde orbits and found that when we vary $a$, the changes in polarization and intensity are minor. Hence, we do not wish to present these results here.

\begin{figure}[h!]
\centering
\includegraphics[width=15cm]{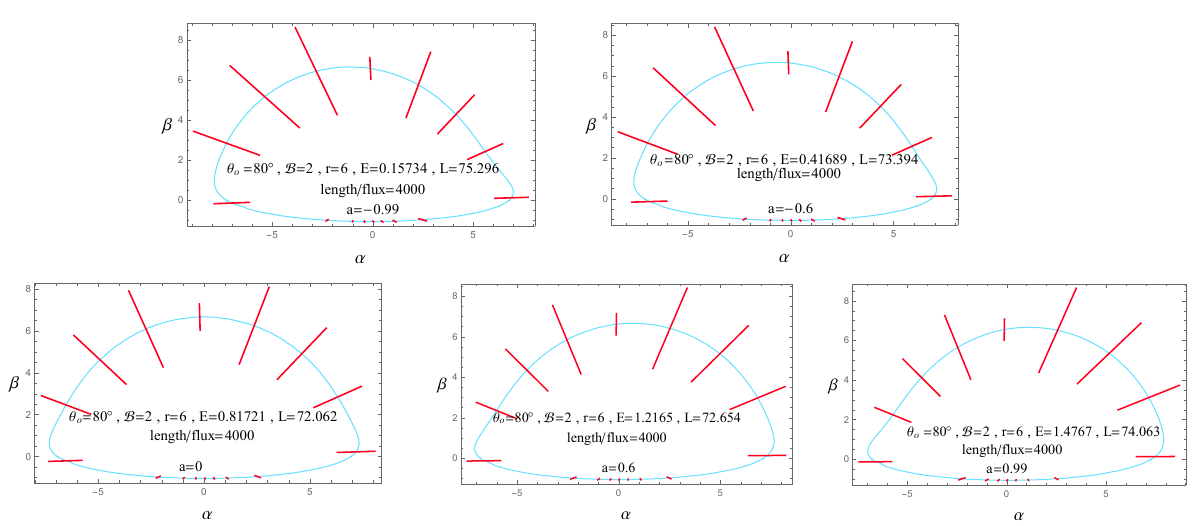}
\caption{Polarized images of charged particles in prograde orbits are produced with $\theta_o=80^\circ$, $\mathcal{B}=2$, and $r_s=6$. We let the spin $a$ vary from $-0.99$ to $0.99$. In this set of plots, we set length/flux = 4000.}\label{figure:80-pro}
\end{figure}

Now we move on to the case that the observational angle is fixed at $\theta_o=80^\circ$, that is, the observer is close to the equatorial plane, and the spin $a$ is varied from $-0.99$ to $0.99$. The images of orbits deviate from circles a lot in this case. The results for the prograde case are shown in Fig. \ref{figure:80-pro}, where we have scaled the length of the polarization flux by $\text{ length/flux}=4000$. From the plot with $a=0$, it seems that the polarizations of both sides centered on $\alpha=0$ are approximately symmetric, which means that the Doppler beaming is not apparent. However, there is a significant difference in the flux between the upper and lower part of the plot, which is caused by the curvature of the strongly curved spacetime.
Meanwhile, at the top of the orbit, where $\alpha=0$, the flux of the polarization is much smaller than those in neighboring places. The reason is that the position $(\alpha=0, \beta=\beta_{max})$ on the screen corresponds to the lights that radiate from the source with the four-momentum perpendicular to the velocity of charged particles in the local frame of the source, and they are much weaker than radiations going in other directions. This phenomenon can usually be observed when $\theta_o$ is bigger. Moreover, the spin of the Kerr black hole also significantly impacts the image. When $a$ is positive, the right side of the image is brighter than the left side and vice versa.

\begin{figure}[h!]
\centering
\includegraphics[width=15cm]{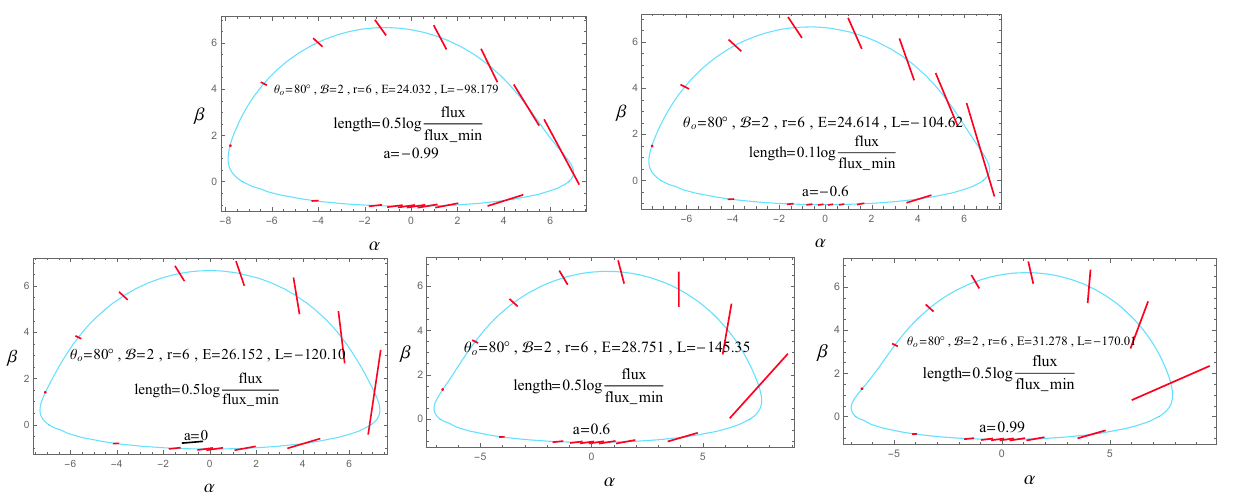}
\caption{Polarized images of charged particles in retrograde orbits are produced with $\theta_o=80^\circ$, $r=6$, and $\mathcal{B}=2$. We also let the spin $a$ vary from $-0.99$ to $0.99$. The red line is scaled by taking the logarithm of flux for the sake of visualization since the Doppler effect is powerful in this case.}\label{figure:80-retro}
\end{figure}

The results for the retrograd case are shown in Figure \ref{figure:80-retro}, where we let $\theta_o=80^\circ$, $r=6$ and $\mathcal{B}=2$ and the spin $a$ go from $-0.99$ to $0.99$. In this case, one of the most obvious phenomena is that the Doppler effect is powerful because the observational angle $\theta_o$ and the energy of the charged particles $E$ are large. Thus for the sake of visualization, we take a logarithm of the flux for the plots. In addition, the right side of the plot, where the particle moves towards the observer, has much stronger radiations. Also, in this case, the black hole spin $a$ has a much larger impact on the polarization. Furthermore, in these plots, we can see a pronounced clockwise rotation of the polarizations as the spin $a$ increases, while this rotation is unapparent for prograde orbits. Besides, we also find that this rotation happens but it's minimal for both prograde and retrograde orbits when $\theta_o$ is small such as $\theta_o=17^\circ$.

It is noteworthy that in actual observations, the detectors or telescopes can only detect electromagnetic radiation of a certain frequency range. Therefore, let us make an estimation of the frequency of  the particle moving on the ISCO . When discussing Kerr black holes, the velocity of the particles on the ISCO increases with the spin parameter $a$. In the case of a near extreme Kerr black hole, the particle velocity approaches the speed of light $c$ on the ISCO, with the radius of the orbit only slightly larger than the black hole mass, that is $r_{ISCO}\sim M$ \cite{Guo:2018kis}. In the International System of Units, we have
\bea
\left(r_{ISCO}\right)_{\text{SI}}=1.5\times10^3\frac{M}{M_\odot}\,.
\eea 
For a solar-mass black hole, the frequency of the particle would be 
\bea
f\sim\frac{c}{2\pi r_{ISCO}}\simeq 30\,\text{kHz} \,
\eea
which maybe at least technically observable. Nonetheless, for the supermassive black holes with a mass of $M\sim 10^9 M_\odot$, the  frequency is roughly $f\sim3\times 10^{-5}$ Hz, which is conceivably unobservable for the ground-based telescope. Hence, it is more likely that the research in the present work can only be applied to the case of black holes of solar mass, from an observational perspective\footnote{We express our utmost gratitude to the anonymous reviewer for drawing our attention to this issue.}.


\section{Summary}\label{summary}

In this work, we studied the circular motions and the polarized images of charged particles in the Kerr spacetime with a weak magnetic field. We focused on the case that the magnetic field and the charge-to-mass ratio are not large, which applies to many charged astrophysical clouds of dust and objects. In this case, the effective potential gets simplified and allows analytical treatment. We found an analytical expression of the radius of the ISCOs in terms of the energy density $E$ and the angular momentum density $L$ and obtained the constraints that $E$ and $L$ should satisfy. It is worth emphasizing that astrophysical objects that widely exist in the universe, which help the condition $\mathcal{B}\ll1$. Thus, our simplified treatment and analytical results are helpful to relevant studies.

Moreover, we revisited how to calculate the polarized images, including the orbits of charged particles, polarization direction, and flux of the synchronize radiations following our previous work \cite{Hu:2022sej} to prepare for numerical computations. We obtained the images of circular orbits through the ray-tracing method. We presented several figures to show the polarized images of charged particles at fixed $\mathcal{B}$ and $r_s$, where $r_s$ was the orbit radius of the source. We would like to stress that in studying the polarized images with the numerical method, we did not assume the approximate condition $\mathcal{B}\ll1$, and our polarized photos were not limited to the case $\mathcal{B}\ll1$.

From our study, we read the influences of the orbit directions, the spins of the black hole, and the observational angles on the polarized images. The images for the prograde and retrograde orbits are very different for the same orbit radius. For example, we found that the so-called Doppler beaming was evident for the retrograde orbits. In addition, we observed a significant influence of the spin $a$ on the polarized images. More precisely, when $a$ increases, the polarization angle rotates clockwise. This phenomenon becomes especially visible for retrograde orbits. The spin $a$ also has a non-negligible effect on the magnitudes of the polarization fluxes. Moreover, we showed the impacts of the observational angle on the polarized images.

It would be illuminating to compare our results with the ones in \cite{Gelles:2021kti}, which also studied the polarized images of equatorial emission in the Kerr geometry. In our study, we focused on a vertical magnetic field corresponding to $(B^{r}=0, B^{\phi}=0, B^{(\theta)})$ in \cite{Gelles:2021kti} somehow, however, our results were closer to the ones in their case $(B^{r}=0, B^{\phi}, B^{(\theta)}=0)$. The main reason is that the electromagnetic radiations in our model are purely from charged particles without self-interactions. At the same time, \cite{Gelles:2021kti} assumed that the synchrotron radiations were from charged fluids, and thus the formulas of the polarization vector of the radiations are different. In addition, the configuration of the magnetic field is chosen in the frame of the source in \cite{Gelles:2021kti}, while we pick the configuration observed by the observers at infinity.

Moreover, the formula Eq. (\ref{pve}) for the polarization vector, which was directly derived from the Maxwell equation in curved spacetime, takes a covariant form and can easily be used in numerical studies. In contrast, the polarization vector in \cite{Gelles:2021kti} is determined in a local frame.

\section*{Acknowledgments}
We would like to thank Y.H. Hou, Y.Shen, H.P. Yan and Z.Y. Zhang for valuable discussions. The work is in part supported by NSFC Grant  No. 12275004, 11775022 and 11873044. MG is also supported by ``the Fundamental Research Funds for the Central Universities'' with Grant No. 2021NTST13.
\appendix

\bibliographystyle{utphys}
\bibliography{citations}

\end{document}